\documentstyle[aps,prl,twocolumn]{revtex}
\begin{document}
\draft 
\def\OP{\raisebox{.2ex}{$\stackrel{\leftrightarrow}{\bf P}$}}
\def\B.#1{{\bbox{#1}}}
\title{The Viscous Lengths in Hydrodynamic
  Turbulence are Anomalous Scaling Functions} 
\author { Victor S.
  L'vov and Itamar Procaccia }
 \address{Department of Chemical
  Physics, The Weizmann Institute of Science, Rehovot 76100, Israel }
\maketitle
 \begin{abstract}
   It is shown that the idea that scaling behavior in turbulence is
   limited by one outer length $L$ and one inner length $\eta$ is
   untenable.  Every n'th order correlation function of velocity
   differences $\bbox{\cal F}_n(\B.R_1,\B.R_2,\dots)$ exhibits its own
   cross-over length $\eta_{n}$ to dissipative behavior as a function
   of, say, $R_1$.  This length depends on $n$ {and on the remaining
     separations} $R_2,R_3,\dots$.  One result of this Letter is that
   when all these separations are of the same order $R$ this length
   scales like $\eta_n(R)\sim \eta (R/L)^{x_n}$ with
   $x_n=(\zeta_n-\zeta_{n+1}+\zeta_3-\zeta_2)/(2-\zeta_2)$, with
   $\zeta_n$ being the scaling exponent of the $n$'th order structure
   function. We derive a class of scaling relations including the
   ``bridge relation" for the scaling exponent of dissipation
   fluctuations $\mu=2-\zeta_6$.
\end{abstract}
The aim of this Letter is to expose the fact that the notion of the
dissipative length in hydrodynamic turbulence is a rich and
interesting concept whose complexity exceeds the expectations of
established models and standard theories \cite{Fri}.  Indeed, during a
few decades the thinking about the universal small scale structure of
turbulence was dominated by Kolmogorov's picture of energy cascade
through an ``inertial interval" which is limited on one side by the
integral scale of turbulence $L$ and on the other side by the
Kolmogorov viscous scale $\eta=(\nu^3/\bar\epsilon)^{1/4}$ where $\nu$
and $\bar\epsilon$ are the fluid's kinematic viscosity and the mean
energy flux in the turbulent flow respectively. During the last decade
there has been a growing concern about the inability of Kolmogorov's
theory to cope with the increasing experimental evidence for
multiscaling (or multifractal) behaviour of higher order structure
functions.  Together with the concern about the statistical theory
there arose a realization that the uniqueness of the viscous length is
suspicious.  Paladin and Vulpiani \cite{87PV}, and also Frisch and
Vergassola \cite{91FV} used the multifractal model of turbulence to
assess the characteristic viscous lengths associated with the higher
order structure functions of velocity differences
\begin{eqnarray}
S_{2n}(R_1) &=& \left<|\B.w(\B.r_1|\B.r'_1)|^{2n} \right> \ , \label{Sn}\\
S_{2n+1}(R_1) &=& \hat\B.R_1\cdot \left<\B.w(\B.r_1|\B.r'_1)|
\B.w(\B.r_1|\B.r'_1)|^{2n} \right> \ , \
\end{eqnarray}
where $\B.w(\B.r_1|\B.r'_1,t)\equiv \B.u (\B.r'_1,t)- \B.u(\B.r_1,t)$
and $\B.u(\B.r,t)$ is the velocity field of the fluid, $\B.R_1\equiv
\B.r'_1-\B.r_1$, and $\hat \B.R_1\equiv \B.R_1/R_1$.  In homogeneous
and locally istoropic turbulence $S_n(R)$ is a function of the
magnitude of $\B.R$, and the viscous length is that value of $R$ at
which the functional dependence of $S_n(R)$ changes from a non-trivial
power law $S_n(R)\sim R^{\zeta_n}$ to a trivial power law that stems
from a Taylor expansion of the velocity differences, $S_n(R)\sim R^n$.
The multifractal model leads to a prediction that this length depend
on the order $n$. In this Letter we argue that a proper discussion of
cross-overs to dissipative behaviour requires the analysis of
functions richer than structure functions. Firstly, we state that the
fundamental object to analyze is the $n$-point correlation of velocity
differences
\begin{equation}
 \bbox{\cal F}_n(\B.r_1,\B.r'_1\dots\B.r_{n},\B.r'_{n}) =
\left<\B.w(\B.r_1|\B.r'_1)\,\dots \B.w(\B.r_{n}
|\B.r'_{n})\right>, \label{Fn}
\end{equation}
which is an $n$-rank tensor. All the separations
$R_i\equiv|\B.r'_{i}-\B.r_{i}|$ and $r_{ij}\equiv|\B.r_i -\B.r_j|$ are
within the "inertial range".  It is generally accepted that this
correlation function is a homogeneous function of its arguments, i.e.
\begin{equation}
\bbox{\cal F}_n(\lambda\B.r_1, \lambda\B.r'_1\dots\lambda\B.r'_{n})
= \lambda^{\zeta_n}\bbox{\cal F}_n(\B.r_1,\B.r'_1\dots\B.r'_{n}). 
\label{assum2}
\end{equation}
It should be understood that quantities like (\ref{Sn}) are obtained
from (\ref{Fn}) by fusing some coordinates together. (In this case all
$\B.r_{ij}\to 0$ and all $\B.R_{i}\to \B.R$). In this process of
fusion one crosses the viscous scale, and it is important to
understand how to do this.

Our discssion will not call for any ad-hoc model of turbulence. It
will be based on two solid building blocks, one being the
Navier-Stokes equations, and the other the fusion rules that were
derived recently. The fusion rules appear naturally in the analytic
theory of Navier-Stokes turbulence \cite{94LL,95LP-b,95LP-d,96LP} and
passive scalar turbulent advection \cite{96LP,95FGLP,95CFKL}, and they
determine the analytic structure of the $n$-order correlation
functions (\ref{Fn}) when a group of coordinates tend towards each
other.  The fusion rules were derived in \cite{96LP} for systems in
which Eq.(\ref{assum2}) holds with universal scaling exponents (i.e
the scaling exponents do not depend on the detailed form of the
driving of the turbulent flows). The fusion rules address the
asymptotic properties of $\bbox{\cal F}_{n}$ when a group (or groups)
of coordinates tend towards a common coordinate withing each group,
while all the other coordinates remain separated by a large distance
$R$.  There are two particular examples of fusion rules that we will
employ in this Letter.  The first pertains to the fusion of one pair
of points. When the distance between one pair is small, $R_1\sim
\rho$, and the separations between all the other coordinates are much
larger, $R_i\sim R$ for $i\ne 1$, then to leading order in $\rho/R$
\begin{equation}
{\cal F}_{n}\propto
 S_{n}(R)S_2(\rho)/ S_2(R). \label{fusion1}
\end{equation}
The second situation pertains to the case in which we have two groups
of fusing coordinates separated by a large distance $R$. When there is
a group of $p$ points separated by a typical distance $\rho_1$, and a
group of $n-p$ points separated by a typical distance $\rho_2$ with a
large distance $R$ between the groups, then
\begin{equation}
{\cal F}_{n}\sim S_{n}(R)
S_p(\rho_1)S_{n-p}(\rho_2)/ S_p(R) S_{n-p}(R). \label{fusion2}
\end{equation}
These forms hold as long as $\rho$, $\rho_1$ and $\rho_2$ are in the
inertial range.

The Navier-Stokes equations for an incompressible velocity field
$\B.u(\B.r,t)$ may be written in the form
\begin{equation}
 \dot{\B.u}({\B.r},t) +\OP\left [{\B.u}({\bf
r},t)\cdot\bbox\nabla
\right]{\B.u}({\B.r},t)=\nu\nabla^2{\B.u}({\B.r},t)\ .
\label{NSpro}
 \end{equation}
 Here $\nu$ is the kinematic viscosity and $\OP$ is the transverse
 projector.  Given the equation of motion we can take the time
 derivative of Eq.(\ref{Fn}). We find \FL
$$
\dot{\bbox{\cal F}}_n=\sum_{j=1}^n
\Big\langle\B.w(\B.r_1|\B.r'_1,t) \dots
\dot\B.w(\B.r_j|\B.r'_j,t)
\dots \B.w(\B.r_n|\B.r'_n,t)\Big\rangle.
$$
Substituting Eq.(\ref{NSpro}), and considering the stationary state
in which $\partial \bbox{\cal F}_n/ \partial t=0$ we find the balance
equations
\begin{equation}
\B.{\cal D}_n(\B.r_1,\B.r'_1;\dots \B.r_n,\B.r'_n)
=\B.{\cal J}_n(\B.r_1,\B.r'_1;\dots
\B.r_n,\B.r'_n)
\ . \label{bal}
\end{equation}
The term $\B.{\cal J}_n$ originates from the viscosity term in (\ref{NSpro}),
\begin{eqnarray}
&&\B.{\cal J}_n(\B.r_1,\B.r'_1;\dots
\B.r_n,\B.r'_n)=\nu\sum_{j=1}^n\left(\nabla_j^2+\nabla_{j'}^2\right)
\langle\B.w(\B.r_1|\B.r'_1)\dots \nonumber \\ &&\dots
\B.w(\B.r_j|\B.r'_j)
\dots \B.w(\B.r_n|\B.r'_n)\rangle \ . \label{Jn}
\end{eqnarray}
The term $\B.{\cal D}_n$ stems from the nonlinear term, and it needs a
bit of algebra to bring to the exact form
\begin{eqnarray}
&&{\cal D}^{\alpha_1\alpha_2\dots \alpha_n}_n(\B.r_1,\B.r'_1;\dots
\B.r_n,\B.r'_n)=\int d\B.r \sum_{j=1}^n  P_{\alpha_j\beta}(\B.r)
\label{Dn}\\
&&\times \langle w_{\alpha_1}(\B.r_1|\B.r'_1)
\dots L^\beta(\B.r_j,\B.r'_j,\B.r)
\dots w_{\alpha_n}(\B.r_n|\B.r'_n) \rangle\ ,
\nonumber\\
&&L^\beta(\B.r_j,\B.r'_j,\B.r)\equiv{1\over n}\sum_{k=1}^n
\big\{ w_\gamma(\B.r_j\!-\!\B.r|\B.r_k) \nabla_j^\gamma\label{L}\\
&&+w_\gamma(\B.r'_j\!-\!\B.r|\B.r'_k)
\nabla_{j'}^\gamma\big\}
w_\beta (\B.r_j-\B.r|\B.r'_j-\B.r).\nonumber
\end{eqnarray}
We are going to argue now that when all the separations $R_j$ are of
the same order of magnitude $R$, the interaction term has a very
simple evaluation, i.e.
\begin{equation}
{\cal D}_n \sim S_{n+1}(R)/R \ . \label{Deval}
\end{equation}
To this aim we need to prove that the integral is local in the sense
that it converges in the ultraviolet and in the infrared.

As the coordinate $\B.r$ is being integrated over, the most dangerous
ultraviolet contribution comes from the region of small $r$. In this
region the projection operator can be evaluated as $1/r^3$. Other
coalescence events of $\B.r$ with other coordinates contribute less
divergent integrands since the projection operator is not becoming
singular. When $r$ becomes small, there are two possiblilites: (i)
$\B.r_j \ne \B.r_k$ and (ii) $\B.r_j =\B.r_k$. In the first case the
correlation function itself is analytic in the region $r \to 0$, and
we can expand it in a Taylor series $Const+ \B.B\cdot\B.r +\dots$.
where $\B.B$ is an $\B.r$-independent vector. The constant term is
annihilated by the projection operator. The term linear in $\B.r$
vanishes under the $d\B.r$ integration due to $\B.r\to -\B.r$
symmetry. The next term which is proportional to $r^2$ is convergent
in the ultraviolet. In the second case we have a velocity difference
across the length $r$. Accordingly we need to use the fusion rule
(\ref{fusion2}), and we learn that the leading contribution is
proportional to $r^{\zeta_2}$. This is sufficient for convergence in
the ultraviolet. We note that the derivative with respect to $r_j$
cannot be evaluated as $1/r$ when $\B.r_j=\B.r_k$. Rather, it is
evaluated as the inverse of the distance between $\B.r_j$ and the
nearest coordinate in the correlation function.
\begin{figure}
\caption{Typical geometry with $(n-1)$ velocity differences in a
 ball of radius$R$ on the left separated by a large distance $r\gg R$
 from a pair  of points on the right.}
\end{figure}

To understand the convergence of $\B.{\cal D}_n$ when the integration
variable $r$ becomes very large we consider the relevant geometry as
shown in Fig.1.  There is one velocity difference across the
coordinates $\B.r_j-\B.r$ and $\B.r'_j-\B.r$ (which is shown on the
right of the figure), $(n-1)$ velocity differences across coordinates
that are all within a ball of radius $R$ (at the left of the figure),
and one velocity difference across the large distance $r$ which is
much larger than $R$. In the notation of this figure the leading order
contribution for large $r$ is obtained from the fusion rules
(\ref{fusion2}) for the situation on the right and (\ref{fusion1}) for
the geometry on the left. The resulting evaluation for the leading
term is $ r^{\zeta_{n+1}}(R_j/ r)^{\zeta_2} (R/r)^{\zeta_{n-1}}$.  On
the face of it, this term is near dangerous. For any anomalous scaling
the integral converges since $\zeta_{n+1}\le \zeta_{n-1}+\zeta_2$ due
to Hoelder inequalities. This convergence seems slow. However, the
situation is in fact much safer.  If we take into account the precise
form of the second-order structure function in the fusion rules we
find that the divergence with respect to $\B.r_j$ translates in fact
to ${\partial S_2^{\beta \gamma}(\B.R_j)/\partial R_{j\gamma}}$ which
is zero due to incompressibility.  The next order term is convergent
even for simple (K41) scaling. This completes the proof of locality of
(\ref{Dn}). The conclusion is that the main contribution to the
integral in (\ref{Dn}) comes from the region $r\sim R$. Therefore the
integral can be evaluated by straightforward power counting leading to
(\ref{Deval}). It should be stressed that a more detailed analysis
demonstrates that when the separations between the coordinates that do
not involve velocity differences, (i.e separations like $r_{jk}$ but
not $R_j$)go to zero, the evaluation does not change.

The evaluation of the quantity $\B.{\cal J}_n$ is more
straightforward. When all the separations $R_j$ and $r_{ij}$ are of
the same order $R$, the correlator in (\ref{Jn}) is evaluated simply
as $S_n(R)$. The Laplacian is then of the order of $1/R^2$. We note
that when $\nu \to 0$ (which is the limit of infinite Reynolds number
Re), this term becomes negligible compared to $D_n$. The ratio ${\cal
  J}_n/{\cal D}_n$ is evaluated as $\nu S_n(R)/RS_{n+1}(R)$, which for
fixed $R$ vanishes in the limit $\nu \to 0$.  Thus the ``balance
equation" becomes a {\em homogeneous} integro-differential equation
$\B.{\cal D}_n =0$ which may have scale-invariant solutions with
anomalous scaling exponents $\zeta_{n+1}\ne (n+1)/3$. It should be
stressed that the evaluation (\ref{Deval}) remains correct for every
term in $\B.{\cal D}_n$, but various terms cancel to give zero in the
homogeneous equation, {\em provided that the scaling exponent
  $\zeta_n$ is chosen correctly}. To make this important point clear
we exemplify it with the simple case $n=2$ for which $\B.D_n$ can be
greatly simplified.  Consider the scalar object ${\cal
  F}_2(\B.r_1,\B.r'_1,\B.r_2, \B.r'_2) \!=\!
\left<\B.w(\B.r_1|\B.r'_1)\cdot \B.w(\B.r_2|\B.r'_2)\right>$. The
terms in the scalar balance equation for this case are exactly
\begin{eqnarray}
{\cal D}_2(\B.r_1,\B.r'_1,\B.r_2,\B.r'_2)&=&d
[S_3(r_{12'})-S_3(r_{12})]/2dr_1\nonumber \\
&+&d[S_3(r_{1'2})-S_3(r_{1'2'})]/2dr'_1
\ , \label{D2}\\
{\cal J}_2(\B.r_1,\B.r'_1,\B.r_{2},\B.r'_{2})
&=&\nu\{\nabla^2_1[S_2(r_{12'})
-S_2(r_{12})]\nonumber \\
&+&\nabla^2_{1'}[S_2(r_{1'2})-S_2(r_{1'2'})]\} \ . \label{J2}
 \end{eqnarray}
 When all the separations are of the order of $R$ we can see
 explicitly that $J_2\sim \nu S_2(R)/R^2$ which is much smaller than
 each term in $D_2$. Considering the scale invariant solution $S_3(R)
 =A R^{\zeta_3}$ where $A$ is a dimensional coefficient, we see that
$$
{\cal D}_2(\B.r_1,\B.r'_1,\B.r_{2},\B.r'_{2})={\zeta_3 A\over 2}\Big[
r_{12'}^{\zeta_3-1}-r_{12}^{\zeta_3-1}+r_{1'2}^{\zeta_3-1}-r_{1'2'}^{\zeta_3
-1}\Big].
$$
Obviously the solution for ${\cal D}_2=0$ requires the unique
choice $\zeta_3=1$ which is the known exponent for $S_3$ \cite{Fri}.
The coefficient $A$ is now determined as $\bar\epsilon$ which is the
mean energy dissipation per unit mass and unit time.

There is a cross-over from the scale invariant solution of the
homogeneous equation to dissipative solutions when ${\cal J}_2$
becomes comparable to any of the terms in ${\cal D}_2$. This happens
when at least one of the separations appearing in (\ref{J2}) becomes
small enough. Denoting the smallest separation as $r_{\rm m}$ we
evaluate ${\cal J}_2\sim \nu S_2(r_{\rm m})/r_{\rm m}^2$. From this we
can estimate, using the balance equation, $S_2(r_{\rm m})\sim
(S_3(R)/\nu R)r_{\rm m}^2\sim \bar \epsilon r_{\rm m}^2/\nu$.  In the
inertial range we have $S_2(r) \sim (\bar\epsilon
r)^{2/3}(r/L)^{\zeta_2-2/3}$.  The viscous scale $\eta_2$ for the
second-order structure function is then determined from finding where
these two expressions are of the same order of magnitude, i.e.
\begin{equation}
\bar\epsilon \eta_2^2/\nu = (\bar\epsilon \eta_2)^{2/3}(r/L)^{\zeta_2-2/3}.
\label{defeta}
\end{equation}
Using the outer velocity scale $U_L$ we estimate $\bar\epsilon \sim
U_L^3/L$ and end
up with
\begin{equation}
\eta_2 \sim L{\rm Re}^{-1/(2-\zeta_2)} \ . \label{eta2}
\end{equation}
Note that this result is not in agreement with the ad-hoc application
of the multifractal model \cite{Fri,87PV,91FV} which predicts $\eta_2
\sim L{\rm Re}^{-2/(2+\zeta_2)}$.

A similar mechanism operates in the general case of $n\ne 2$. As long
as all the separations are in the inertial interval $\B.{\cal J}_n$ is
negligible.  When one separation e.g. $r_{12}$ diminishes towards
zero, and all the other separations are of the order of $R$, the
internal cancellations leading to the homogenous equation $\B.{\cal
  D}_n=0$ disappear, and $\B.{\cal D}_n$ is evaluated as in
(\ref{Deval}).  The term $\B.{\cal J}_n$ is now dominated by one
contribution that can be written in short-hand notation as
$\nu\nabla^2_1F_n(r_{12},\{R\})$. We can solve for $F_n(r_{12},\{R\})$
in this limit:
\begin{equation}
F_n(r_{12},\{R\})\approx r_{12}^2 S_{n+1}(R)/ \nu R \ .\label{dissFn}
\end{equation}
On the other hand we have, from the fusion rule (\ref{fusion2}), the
form of the same quantity when $r_{12}$ is still in the inertial
range, i.e.  $F_n(r_{12},\{R\}) \approx S_2(r_{12})S_n(R)/S_2(R)$. To
estimate the viscous scale $\eta_n$ we find when these two evaluations
are of the same order.  The answer is
\begin{equation}
\eta_n(R) = \eta_2 \Big({R\over L}\Big)^{x_n},\quad
x_n={\zeta_n+\zeta_3-\zeta_{n+1}-\zeta_2\over
2-\zeta_2} \ . \label{finaletan}
\end{equation}
We note that the Hoelder inequalities guarantee that $x_n>0$ and
increases with $n$. We see that the viscous ``length" is actually an
anomalous scaling function.

Next we show that in the same spirit we can derive important (and
exact) scaling relations between the exponents $\zeta_n$ of the
structure functions and exponents involving correlations of the
dissipation field.  We consider correlations of the type \FL
\begin{eqnarray}
&&\B.{\cal K}_{\epsilon}^{(n)}\equiv \left<\epsilon(\B.r_1)
\B.w(\B.r_2|\B.r'_2)\!\dots \!\B.w(\B.r_n|\B.r'_n)\right>\propto
R^{-\mu^{(1)}_n},
\label{Ken}\\
&&\B.{\cal K}_{\epsilon\epsilon}^{(n)}\equiv
\left<\epsilon(\B.r_1)\epsilon(\B.r'_1)
\B.w(\B.r_2|\B.r'_2)\!\dots \!\B.w(\B.r_n|\B.r'_n)\right>\propto
R^{-\mu^{(2)}_n},
\label{Keen}
\end{eqnarray}
where $R$ is a typical separation between any pair and
$\epsilon(\B.r)\equiv \nu|\nabla \B.u(\B.r)|^2$, and we are interested
in the scaling relations between the exponents $\mu_n$ and the
exponents $\zeta_n$. Note that $\mu^{(2)}_0$ in this notation is the
well studied \cite{93Pra,93SK}exponent of dissipation fluctuation
which is denoted as $\mu$. This relation is almost at hand for
$\mu^{(1)}_n$. We see this by writing
\begin{equation}
\B.{\cal K}_{\epsilon}^{(n)}=\nu\lim_{r_{12}\to 0}\B.\nabla_1
\B.\nabla_2\B.{\cal F}_n(r_{12},\{R\}).\label{KF}
\end{equation}
Using the result (\ref{dissFn}) we find immediately
\begin{equation}
\mu^{(1)}_n = 1-\zeta_{n+3} . \label{mu1n}
\end{equation}
The scaling relations satisfied by $\mu^{(2)}_n$ require
considerations of the second time derivative of the correlation
(\ref{Fn}).  \FL
\begin{eqnarray}
\ddot{\bbox{\cal F}}_n=\sum_{i,j=1}^n
\langle\B.w(\B.r_1|\B.r'_1,t) \dots\dot\B.w(\B.r_i|\B.r'_i,t)
\nonumber\\
\dots\dot\B.w(\B.r_j|\B.r'_j,t)
\dots \B.w(\B.r_n|\B.r'_n,t)\rangle.
\end{eqnarray}
Using the Navier-Stokes equations for the time derivatives we derive a
new balance equation $\B.{\cal D}^{(2)}_n+\B.{\cal B}^{(2)}_n=\B.{\cal
  J}^{(2)}_n$ where, using the definition (\ref{L}),
\begin{eqnarray}
&&\B.{\cal D}_n^{(2)}=\int d\B.r d\B.r' \sum_{i,j=1}^n  \B.P(\B.r)\B.P(\B.r')
 \Big\langle \B.w(\B.r_1|\B.r'_1)\label{D2n}\\
&&\dots \B.L(\B.r_i,\B.r'_i,\B.r)\dots \B.L(\B.r_j,\B.r'_j,\B.r')
\dots \B.w(\B.r_n|\B.r'_n)\Big \rangle \ .\nonumber
\end{eqnarray}
Using the fusion rules and following steps similar to those described
above, we can prove that the integrals over $\B.r$ and $\B.r'$
converge. Accordingly, when all the separations are of the order of
$R$, every term in $D^{(2)}_n$ is evaluated as $S_{n+2}(R)/R^2$.  The
term $\B.{\cal J}^{(2)}_n$ takes on the form
\begin{eqnarray}
&&\B.{\cal J}^{(2)}_n=\nu^2\sum_{i,j=1}^n\left(\nabla_i^2+\nabla_{i'}^2\right)
\left(\nabla_j^2+\nabla_{j'}^2\right)\label{J2n}\\
&\times&\langle\B.w(\B.r_1|\B.r'_1)\dots \B.w(\B.r_i|\B.r'_i)\dots
\B.w(\B.r_j|\B.r'_j)
\dots \B.w(\B.r_n|\B.r'_n)\rangle \ .\nonumber
\end{eqnarray}
As before, when all the separation in this quantity are of the order
of $R$, the Laplacian operators introduce factor of $1/R^2$ and the
evaluation of this quantity is $\B.{\cal J}^{(2)}_n\sim
\nu^2S_n(R)/R^4$. Clearly this is negligible compared to typical terms
in $D^{(2)}_n$. The quantity $\B.{\cal B}^{(2)}_n$ contains a cross
contribution with one Laplacian operator and one nonlinear term with a
projection operator. The integral is again local, and one can show
that the evaluation is $\B.{\cal B}^{(2)}_n\sim \nu S_{n+1}(R)/R^3$
which is also negligible compared to typical terms in $D^{(2)}_n$.

Now we consider the fusion of two pairs of coordinate, e.g. $r_{12}\to
0$ and $r_{34}\to 0$. As before, the cancellations in $\B.{\cal
  D}^{(2)}_n$ are eliminated, and the evaluation of a typical term
becomes the evaluation of the quantity.  The other two terms in the
balance equation also become of the same order because the Laplacian
operators $\nabla_1^2$ and $\nabla_3^2$ are evaluated as $r_{12}^{-2}$
and $r_{34}^{-2}$ respectively. As before we can consider the
resulting balance equation as a differential equation for
$F_n(r_{12},r_{34},\{R\})$. The leading term in this equation is
$$
4\nu^2\nabla_1^2  \nabla_2^2 F_n(r_{12},r_{34},\{R\})
\approx \B.{\cal B}^{(2)}_n
+\B.{\cal D}^{(2)}_n \sim S_{n+2}(R)/R^2 .
$$
The solution is
\begin{equation}
{\cal F}_n(r_{12},r_{34},\{R\})\sim r_{12}^2  r_{34}^2S_{n+2}(R)/ \nu^2 R^2 \ .
\label{dissF2n}
\end{equation}
Finally we can write the quantities $\B.{\cal
  K}_{\epsilon\epsilon}^{(n)}$ in terms of the correlation function as
\begin{equation}
\B.{\cal K}_{\epsilon\epsilon}^{(n)}=\nu^2\lim_{r_{12},r_{34}\to 0}\B.\nabla_1
\B.\nabla_2\B.\nabla_3\B.\nabla_4\B.{\cal F}_n(r_{12},r_{34},\{R\}).\label{KF1}
\end{equation}
Using (\ref{dissF2n}) here we end up with the evaluation
\begin{equation}
\B.{\cal K}_{\epsilon\epsilon}^{(n)}\sim S_{n+6}/R^2\propto
R^{-\mu^{(2)}_n},\quad
\mu^{(2)}_n=2-\zeta_{n+6} . \label{mun}
\end{equation}
For the standard exponent $\mu=\mu_0^{(2)}$ we choose $n=0$ and obtain
the phenomenologically proposed ``bridge relation" $\mu=2-\zeta_6$. To
our best knowledege this is the first solid derivation of this scaling
relation. In general, if we have $p$ dissipation fields correlated
with $n$ velocity differences the scaling exponent can be found by
considering $p$ time derivatives of (\ref{Fn}), with the final result
\begin{equation}
\mu^{(p)}_n=p-\zeta_{n+3p}.\label{mup}
\end{equation}
We see that Eqs.(\ref{mu1n}), (\ref{mun}) and (\ref{mup}) can be
guessed if we assert that {\em for the sake of scaling purposes} the
dissipation field $\epsilon(\B.r)$ can be swapped in the correlation
function with $w^3(\B.r_1|\B.r'_1)/R_1$, where $R_1$ is the
characteristic scale. This reminds one of the Kolmogorov refined
similarity {\em hypothesis}. We should stress that (i) our result does
not depend on any uncontrolled hypothesis, and (ii) it does not imply
the correctness of the hypothesis. Our result is implied by the
refined similarity hypothesis, but not vice versa.

\noindent
{\bf Acknowledgments.} This work was supported in part by the German
Israeli Foundation, the Minerva Center for Nonlinear Physics, and the
Naftali and Anna Backenroth-Bronicki Fund for Research in Chaos and
Complexity.


\begin{references}
\bibitem{Fri}
U.Frisch {\em Turbulence} (Cambridge 1995).

\bibitem{87PV}
G. Paladin and A. Vulpiani, Phys.Rev.A {\bf 35}, 1971 (1987).

\bibitem{91FV}
U. Frisch and M. Vergassola, Europhys. Lett. {\bf 14}, 439 (1991).

\bibitem{94LL}
V.V. Lebedev and V.S. L'vov, Pis'ma Zh. Eksp. Teor.
 Fiz. {\bf 59}, 546 (1994) [JETP {\bf 59}, 577 (1994)].

\bibitem{95LP-b}
V.S. L'vov and I.~Procaccia, Phys.~Rev.~E, {\bf 52}, 3858 (1995).

\bibitem{95LP-d}
V.S. L'vov and I.~Procaccia, Phys.~Rev.~E {\bf 53}, 3468, (1996).

\bibitem{96LP}
V.S. L'vov and I. Procaccia, Phys. Rev Lett.{\bf 76}, 2896 (1996).

\bibitem{95FGLP}
A.L. Fairhall, O. Gat, V.S. L'vov and I. Procaccia Phys. Rev. E,
  {\bf 53}, 3518 (1996).

\bibitem{95CFKL}
M. Chertkov, G. Falkovich, I. Kolokolov
and V. Lebedev, Phys. Rev. E {\bf 52}, 4924 (1995).

\bibitem{93Pra}
A.A. Praskovsky, Phys. Fluids A{\bf 4}, 2589 (1992).

\bibitem{93SK}
K.R. Sreenivasan and K.R. Kailasnath, Phys. Fluids A{\bf 5}, 512 (1993).

\end{references}
\end{document}